\documentclass{elsart}

\usepackage{epsfig}
\usepackage{amssymb,amsmath}  % Les maths au standard de l'AMS
\usepackage{natbib}

\newcommand{\eg}{{\em e.g., }}           % e.g.
\newcommand{\ie}{{\em i.e., }}           % e.g.
\def\be{\begin{equation}}
\def\ee{\end{equation}}
\def\ba{\begin{array}}
\def\ea{\end{array}}

\begin{document}
\begin{frontmatter}
\title{Material-independent crack arrest statistics}
\author[FH]{Yann {\sc Charles,}}
\author[SR]{Damien {\sc Vandembroucq,}}
\author[FH]{Fran\c{c}ois {\sc Hild}} and
\author[SR]{St\'{e}phane {\sc Roux}\thanksref{COR}}
\thanks[COR]{to whom correspondence should be addressed.\\
e-mail:~{\tt stephane.roux@saint-gobain.com}}
\address[FH]{LMT-Cachan\\
ENS de Cachan / CNRS-UMR 8535 / Universit\'{e} Paris 6\\
61 avenue du Pr\'{e}sident Wilson, F-94235 Cachan Cedex, France.\\}
\address[SR]{Surface du Verre et Interfaces,\\
Unit\'{e} Mixte de Recherche CNRS/Saint-Gobain\\
39 quai Lucien Lefranc, F-93303 Aubervilliers Cedex, France.\\}
\begin{abstract}
The propagation of (planar) cracks in a heterogeneous brittle
material characterized by a random field of toughness is
considered, taking into account explicitly the effect of the crack
front roughness on the local stress intensity factor.  In the
so-called strong-pinning regime, the onset of crack propagation
appears to map onto a second-order phase transition characterized
by universal critical exponents which are independent of the local
characteristics of the medium.  Propagation over large distances
can be described by using a simple one-dimensional description,
with a correlation length and an effective macroscopic toughness
distribution that scale in a non-trivial fashion with the crack
front length.  As an application of the above concepts, the arrest
of indentation cracks is addressed, and the analytical expression
for the statistical distribution of the crack radius at arrest is
derived.  The analysis of indentation crack radii on alumina is
shown to obey the predicted algebraic expression for the radius
distribution and its dependence on the indentation load.
\end{abstract}
\end{frontmatter}

\newpage
%
%################################################################
%################################################################
%
\section{Introduction}

Brittle materials, such as ceramics and glasses, are known to be
extremely sensitive to bulk and surface defects, from which cracks
can be initiated eventually leading to failure.  This extreme
sensitivity calls for a statistical analysis of crack initiation,
which has been extensively developed following the pioneering work
of Weibull~\citep{47}.  In this ``weakest-link''
approach~\citep{17}, the analysis is focused on the initiation
stage, implicitly assuming that the propagation stage is obtained
systematically over unlimited distances, due to the lower loading
needed for this stage as compared to the inception of propagation.

However, in some cases, even though cracks are initiated, they
will not propagate over large distances, and their presence may
still be acceptable in service condition of a given structure. For
example, in ceramic~/~metal assemblies, residual stresses caused
by the coefficient of thermal expansion mismatch may prevent
cracks to traverse the brittle part so that a weakest link
hypothesis does not apply~\citep{1252}. In brittle-matrix
composites, crack arrest is also observed due to the bridging
forces induced by the fibers~\citep{98}. Other examples of such
confined cracks are those produced by indentation~\citep{1431}. In
the latter case, the stress field has a rapid decay with distance
from the indentation point, and hence a crack that can easily be
nucleated may stop shortly after initiation.

The mechanical treatment of crack arrest in a brittle material is
standard: namely, Linear Elastic Fracture Mechanics generally
holds without much restriction, and hence, a crack is expected to
be arrested as soon as the stress intensity factor $K$ becomes
smaller than the toughness $K_c$.  The purpose of the present
paper is to investigate the effect of a statistical distribution
of toughness. A statistical model is considered for the steady
propagation of an extended planar crack front through a random
landscape of toughness.  This situation is shown to give rise to a
genuine phase transition at the onset of propagation. The
recognition of this critical character opens to numerous
interesting properties, not only concerning the complexity of the
spatio-temporal development of crack propagation, but also and
more importantly on the statistical distribution of the global
stress intensity factor characterizing the entire front.  In
particular, {\em universality} justifies the {\em independence} of
some properties on the details of the random toughness field. At a
larger scale, the problem of crack arrest is considered in a
non-uniform loading geometry. In this situation, some of the
features revealed by the previous analysis can be used.  As an
application of the above concepts, the case of indentation cracks
is considered since they are naturally arrested because of the
decaying stress intensity factor.  The semi-circular geometry of
the crack front will allow for an explicit analysis of the
statistical distribution of crack arrest distance. The results are
applied on indentation of sintered alumina samples. Four different
statistical records of indentation crack length for different
loads on the same material are tested against the theoretical
predictions.

\section{Crack pinning}

The three-dimensional problem where the crack front may be pinned
at different locations leads to a complex problem, because elastic
interactions affect the local stress intensity factor along the
crack front.  As a result, the local stress intensity factor
depends on the overall front geometry in a non-local
manner~\citep{1426,1427}. Curtin~\citep{1578,1579} performed a
simplified analysis of a similar problem. A modelling of planar
crack propagation in a random environment was also proposed by
Schmittbuhl {\it et al.}~\citep{1600}, Ramanathan and
Fisher~\citep{1591} and Bouchaud \etal ~\citep{1573} that revealed
a complex spatio-temporal organization of the propagation, and
scale-invariant crack front morphology. These studies show that
the macroscopic toughness is not simply given by the average
toughness and that tougher elements impede crack growth in a
lesser extend than with a strongest link assumption due to the
stress intensity factor enhancement caused by crack front
curvature. One of the most striking predictions of these
approaches is the scale invariant nature of the crack front or
crack surface roughness, which is qualitatively (unfortunately not
quantitatively) in agreement with a large body of experimental
analysis, a review of which is presented in \citep{15731}.

\begin{figure}
 \centerline{\epsfxsize=.9\hsize \epsffile{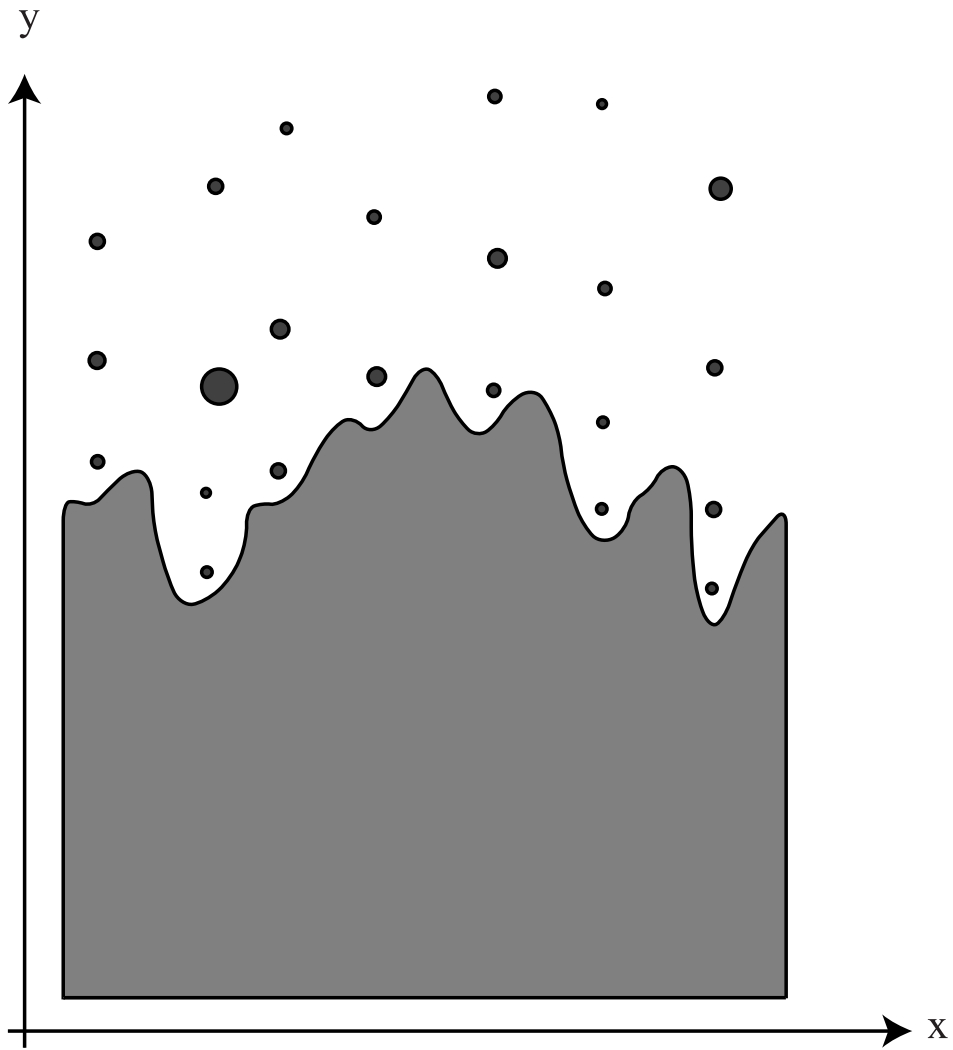}}
 \caption{\label{fi:crack}
 Depiction of crack propagation along the $y$-direction
 through a random toughness field shown here as a collection
 of inclusions which may pin the crack front.}
\end{figure}

This problem, at this level of generality, is extremely complex,
and simplifying assumptions are proposed to address it. First, the
crack remains planar in the $(x,y)$-plane where the mean crack
front is aligned along the $x$-direction, whereas the $y$-axis is
the direction of crack propagation (Fig.~\ref{fi:crack}). Second,
the distribution of local toughness is sufficiently narrow so that
a first-order perturbative analysis can be used to characterize
the coupling of the local stress intensity factor $K(x)$ to the
overall front morphology $h(x)$. Third, the elastic properties of
the medium can be considered as homogeneous. Those assumptions
allow one to use a result derived by Gao and Rice~\citep{1583} that
relates the local stress intensity factor $K(x,h(x))$ along the
crack front to the crack front shape $h(x)$, and the reference
stress intensity factor, $K_0$, which would result from the same
loading with a straight crack front at the same mean position $h_0
= \langle h \rangle_x$
\begin{equation}
    \label{eq:perturb}
    K(x,h(x))= K_0 \left( 1 + \frac{1}{\pi} \int
    \frac{h(x)-h(x')}{(x-x')^2} \mathrm{d}x' \right).
\end{equation}
As discussed in \citep{1611}, those assumptions are not yet
sufficient to specify entirely the problem at hand.  The
correlation function of the toughness pattern comes into play to
discriminate two regimes:

\begin{itemize}

\item
for large correlation lengths along the propagation direction,
a weak-pinning regime is obtained, where the crack front is only
mildly affected by the disorder and hence the crack advance at
each point along the front is continuous as the mean position of
the crack front $\langle h \rangle_x$ moves;

\item
for small correlation lengths along the propagation direction, the
front undergoes a series of jumps from one stable position to the
next.  This regime, referred to as ``strong-pinning,'' is the most
interesting in the sense that the effective toughness is expected
to depart from the arithmetic average of the local toughness (in
contrast to the weak pinning regime).

\end{itemize}
In the following, only the latter regime is considered with a
specific design of the local toughness distribution so that
strong-pinning will be guaranteed.

\section{Propagation model}

Let us consider a planar crack front extending over the $x$-axis
and propagating along the $y$-direction.  The geometry of the
front is described by $h(x,t)$, the $y$-coordinate of the front at
an abscissa $x$ and time $t$.  The model describes the propagation
of the crack from one pinned configuration to the next.  A time
coordinate is introduced to label the sequence of crack front
geometries.  However, as explained below, no physical
time-dependent propagation rule is considered, and hence this
label is used as a convenient way of characterizing the mean crack
advance. The front is discretized along the $x$-axis with a
constant interval whose physical meaning can be interpreted as the
correlation length of the toughness along the $x$-direction. The
assumption that the crack front can be represented as a
single-valued function of $x$ is a necessary assumption for the
use of  the perturbative result~(\ref{eq:perturb}). The latter
imposes the slope of the front to be small compared to
unity~\citep{1583}. This in turn has consequences on the amplitude
of the local toughness distribution, which has to be small
compared to the mean value, $\overline K_c$. This quantity is the
small parameter used in all subsequent expansions.

Under an external load that would produce a stress intensity
factor $K_0(t)$ for a straight front, Eq.~(\ref{eq:perturb})
allows one to estimate the local stress intensity factor,
$K(x,t)$. The local toughness along the crack front
$K_c(x,h(x,t))$ is assumed to be distributed without spatial
correlation from a given distribution $p(K_c)$. To make the
problem non-dimensional, the scaled stress intensity factor is
defined as $k=(K-{\overline K_c})/{\overline K_c}$, and the local
dimensionless toughness $k_c=(K_c-{\overline K_c})/{\overline
K_c}$. The local stress intensity factor can thus be written, as a
first order approximation,
\be
 k(x,t)=k_0(t)+k_{el}(x,t),
\ee
with
\be
 k_{el}(x,t)=\frac{1}{\pi}\int \frac{h(x,t)-h(x',t)}{(x-x')^2}
 \mathrm{d}x'.
\ee
One can compute the maximum level for which the crack would not
propagate, $k_0(t)<k_c(x,h(x,t))-k_{el}(x,t)$, for all $x$ along
the front.  Thus, the onset of crack propagation is given by a
level such that the scaled global stress intensity factor assumes
the value
\be
 \label{eq:critloadt}
 \kappa(t)=\min_x [k_c(x,h(x,t))-k_{el}(x,t)].
 \ee
For such a level, at one point $x_{act}(t)$ along the front the
stress intensity factor matches exactly the local toughness. Let
us now introduce the way the crack propagation is modeled. At the
point $x_{act}(t)$, the crack can overcome the local toughness and
advance.  It will directly jump to a new obstacle
$h(x_{act}(t),t+1)= h(x_{act}(t),t)+\delta h$, where $\delta h$
can be either a constant length, or a random one. It characterizes
the correlation along the propagation direction, and there lies
the ``strong-pinning'' nature of the present model. At this new
position, another local toughness is encountered, which is {\em
uncorrelated} with the previous ones.  All other sites are
supposed to remain pinned during this elementary move of the crack
front. The new front geometry and local toughness distribution is
treated the same way.  This means that the external loading is
adjusted in time so that one constantly remains at the onset of
propagation given by Eq.~(\ref{eq:critloadt}), irrespective of the
variation of $\kappa(t)$.  The physical time needed to achieve the
elementary move of the crack front is not introduced here, and
hence time appears here as a discrete label given to the different
crack configurations.  However, when inertia effects are
negligible, one can reconstruct from the above signal $\kappa(t)$
what would be the evolution of the system under any specific
history of loading in a quasi-static regime.  In particular, for a
subcritical but constant loading, $k_0$, starting from any
configuration, one can compute the expected crack advance before
reaching a pinned configuration for which $\kappa(t)>k_0$ for the
first time.

For a crack front having a finite extent (with free or periodic
boundary conditions), one can follow numerically the change of a
crack front for arbitrarily long times, and record statistical
information concerning either the crack front geometry, the
activity of the front with time [\ie position of $x_{act}(t)$], or
the loading $\kappa(t)$.  The latter signal is of special interest
since it provides an estimate of the (reduced) critical loading
$\kappa^*$ for which an unlimited propagation would be obtained
\be
 \kappa^*=\max_t[\kappa(t)],
\ee
and thus the latter corresponds to the macroscopic effective
toughness of the medium, for extended propagation, while
$\kappa(t)$ can be understood as an instantaneous toughness taking
into account interactions between local toughness values through
the dependence in the crack front geometry.  As shown below, the
crack propagation can be compared to a phase transition from the
pinned state to an unpinned (\ie propagation) regime, where the
loading plays the role of the control parameter.  As usually
observed in such second-order phase transition, the behavior close
to $\kappa^*$ can be characterized by a series of critical
exponents which are universal, \ie independent of the local
disorder introduced in the model, such as the distribution
$p(K_c)$ of local toughness (see \citep{1615} for a presentation
of the background and a review of other occurrences of criticality
in a variety of applications including mechanical and geophysical
examples).

\section{Statistical properties of the steady state}

After a transient regime influenced by the initial crack geometry,
the model gives rise to a steady state during which a number of
statistical observables can be defined and characterized. The main
properties of this steady state are recalled.  More details can be
found in \citep{1596,1595}.

\subsection{Front roughness}

The first interesting observation concerns the crack front
morphology.  No specific scale of roughness is evidenced between
the discretization scale and the system size, but rather a scale
invariant geometry appears.  The crack front is observed to be a
self-affine object such that the r.m.s. height difference between
two points distant of $\Delta x$ scales as
\be
 \langle [h(x+\Delta x)-h(x)]^2\rangle_x^{1/2}\propto \Delta
 x^\zeta,
\ee
with $\zeta\approx 0.37$, {\em irrespective} of the disorder.  The
same property can be analyzed with different tools~\citep{1614},
such as Fourier power spectra, wavelet analysis, probability
density function (p.d.f.) of height differences, all revealing the
same property.

\subsection{Activity}

The spatio-temporal distribution of activity [\ie map of
$x_{act}(t)$ vs.\ $t$] displays a highly non-trivial organization,
without any characteristic scale, neither in space nor in time.
More precisely, by studying the p.d.f. of the difference
$d=|x_{act}(t+\Delta t)-x_{act}(t)|$ at a fixed time interval, two
regimes are found:

\begin{itemize}
\item
at a small distance, the p.d.f. is uniform;

\item
at large distances, the p.d.f. decays as a power-law $d^{-2}$ that
reflects the kernel of the interaction term in
Eq.~(\ref{eq:perturb}).

\end{itemize}
The cross-over distance between these two regimes, $d^*$ depends
on the time interval $\Delta t$ as a power-law
\be
  d^*\propto \Delta t^{1/z},
\ee
where $z$ is the so-called dynamic exponent, related to the
roughness exponent $\zeta$ by
\be
 z=1+\zeta.
\ee
Such a behavior can be recovered in a class of stochastic
processes, \ie Linear Fractional Stable Motion~\citep{1616} that
contains both a wide distribution of elementary $d$, and
long-range temporal correlations.

The above property reveals that the propagation consists in
localized bursts of activity, where the clustering of activity in
space and time can be decomposed hierarchically from the larger
ones down to the smaller ones in a scale-invariant fashion.

\subsection{Distribution of effective toughness}
\label{se:keff}

The two above properties dealt with geometric organization that
may be difficult to access experimentally. However, the model also
provides access to the loading, and hence the effective
macroscopic toughness either at each instant of time $\kappa(t)$
or globally $\kappa^*$.   The interesting feature is that the
signal $\kappa(t)$ contains some information about the
spatio-temporal organization of  activity.

\begin{figure}
\centerline{\epsfxsize=.9\hsize \epsffile{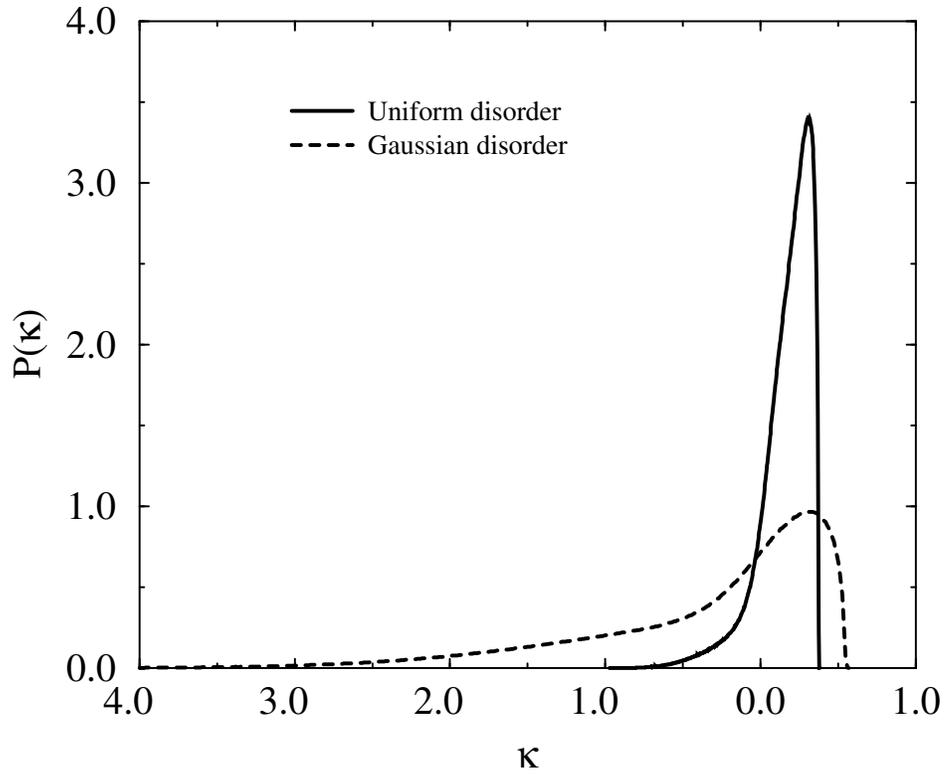}}
\caption{\label{fi:fc}Distribution of the effective toughness
$\kappa(t)$ obtained for a uniform toughness distribution in
$[0;1]$ (solid line) and centered reduced Gaussian (dashed line).}
\end{figure}

The first observation is the statistical distribution of $\kappa
(t)$ which shows a long tail of low effective toughness, but an
abrupt fall-off of the distribution near the maximum value
$\kappa^*$.  The latter can be characterized to behave singularly
as
\be
  \label{eq:critload}
  P(\kappa)\propto (\kappa^*-\kappa)^\beta,
\ee
{\em irrespective} of the microscopic toughness distribution $p(K_c)$.
Figure~\ref{fi:fc} shows the distributions $P(\kappa)$ for two
different distributions $p(K_c)$, either uniform in $[0;1]$ or
centered reduced Gaussian.

\begin{figure}
\centerline{\epsfxsize=.9\hsize \epsffile{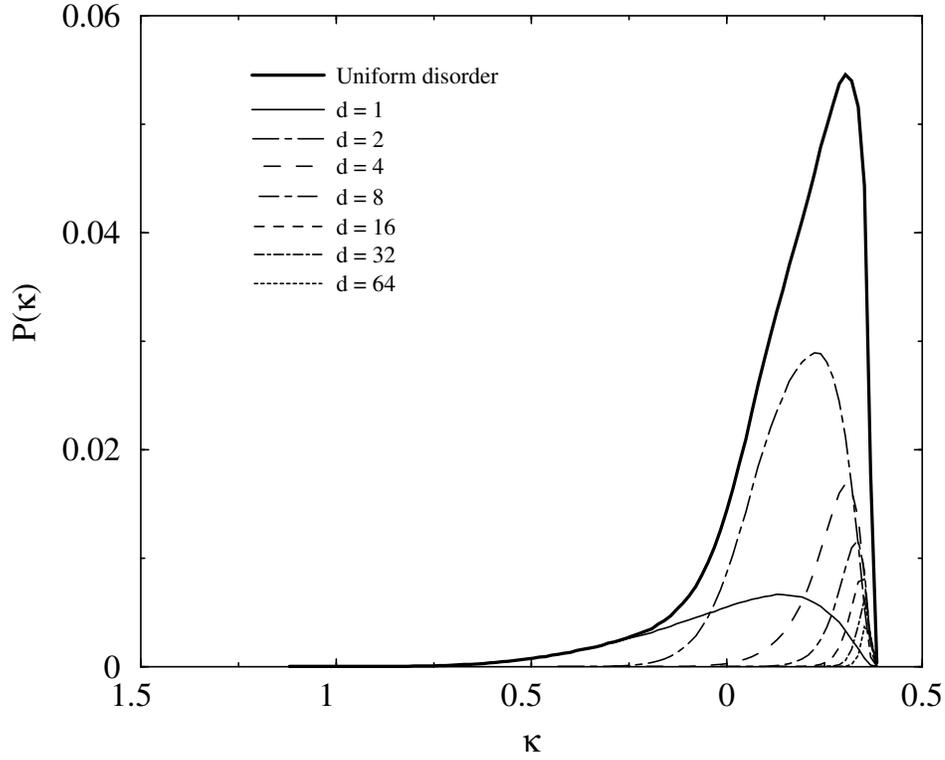}}
\caption{\label{fi:fc2}Distribution of the effective toughness
$\kappa(t)$ obtained for a uniform toughness distribution in
$[0;1]$, and conditioned distributions for geometrically
distributed jump distances from $d=1$ to $d=64$ progressively
concentrating close to $\kappa^*$. The conditioned distributions
are not normalized so that they represent their contribution to
the overall distribution $P(\kappa)$.}
\end{figure}

To make the connection with the activity map, it is instructive to
study the distribution of effective toughness $P(\kappa(t),d)$
conditioned by a jump between the activity site
$|x_{act}(t+1)-x_{act}(t)|=d$. The clustered nature of the
activity suggests that one needs to reach a locally strongly
pinned configuration to observe a large distance jump of the
active site.  Thus the distribution $P(\kappa(t),d)$ is expected
to concentrate progressively closer to $\kappa^*$ as $d$
increases~\citep{1595}. This is what is observed in
Fig.~\ref{fi:fc2}. Moreover, as $d$ becomes large compared to the
discretization scale, it can be seen that the distribution
$P(\kappa,d)$ converges to a unique form up to a scaling factor.
Hence, one can write
\be
 \label{eq:loaddis_d}
 P(\kappa,d)=d^\alpha\psi[(\kappa^*-\kappa)d^{\alpha}],
\ee
so that a characteristic correlation length $\ell_\parallel$ can
be derived for each loading $\ell_\parallel \propto
(\kappa^*-\kappa)^{-\nu}$ where $\nu=1/\alpha$.  The physical
meaning of such a length scale is that for a subcritical loading
$k_0=\kappa$, starting from any configuration in the steady state,
the activity is localized in a region of extent $\ell_\parallel$.
The time needed to reach a stable configuration can be obtained
from the previous scaling properties, $\Delta t \propto
\ell_\parallel^z \propto (\kappa^*-\kappa)^{-\nu z}$.  It can be
demonstrated~\citep{1582,1617} that
\be
  \nu=\frac{1}{1-\zeta},
\ee
or this scaling property can be read off from the interaction
kernel in Eq.~(\ref{eq:perturb}).

Knowing the statistics of jumps $d$ distributed as a power-law of
exponent $-2$, together with the scaling properties of the
conditioned distribution $P(\kappa,d)$, allows one to compute the
overall distribution of $\kappa$ in the vicinity of $\kappa^*$. It
provides the critical behavior described by
Eq.~(\ref{eq:critload}) with
\be
    \beta=\nu-1=\frac{\zeta}{1-\zeta}.
\ee
Table~\ref{tab:expo} recalls the theoretical expressions of all
the critical exponents introduced so far, together with their
numerical value.  All exponents can be expressed as a function of
one of them (\eg $\zeta$).  However, the numerical value of the
latter is not known theoretically. Numerical simulations are
needed to obtain an accurate estimate. Functional renormalization
group methods have been used~\citep{1618,1619} to provide an
analytic route to estimate $\zeta$. However, it remains a
perturbative method, which even when computed up to second
order~\citep{1620}, remains approximative.

\section{Crack arrest}

The discussion has been focused here on the steady state
propagation in a homogeneous environment.  With those results, the
problem of crack arrest can be addressed in a geometry where the
stress intensity factor decreases with the propagation distance.

In \citep{1577,1272,1252,1322}, the problem is treated by using a
simple one dimensional description.  Only the mean propagation
distance $\langle h\rangle_x$ is used to parameterize the crack
advance. Accordingly, the stress intensity profile, $K(\langle
h\rangle_x)$, is introduced, together with some effective
toughness $K_c(\langle h\rangle_x)$.  At this level of
description, the effective toughness is assumed to be given by a
random function, with a correlation length representative of the
microstructure. Thus this random function is modeled as a
piecewise constant function over a ``grain size'' $1/\lambda$,
without any correlation above this scale.  The grain size is
either constant~\citep{1252,1322} or described by using a Poisson
mosaic~\citep{1272}. The p.d.f. for the effective toughness,
$f(K_c)$ is introduced.

Let us briefly recall the argument developed in \citep{1252,1322}.
Let $F(K)=P(K_c<K)$ be the cumulative probability of effective
toughness $F(K)=\int_0^{K} f(K_c) \mathrm{d}K_c$. The probability
$Q(h)$ that the crack arrest distance be larger than $h$ is then
computed by writing that the condition $K_c(y)<K(y)$ hold for all
$y \le h$ (\ie a ``strongest'' link assumption)
\be
  \label{eq:1Dd}
  Q(h) = \prod_{i=1}^{h\lambda} F(K(i/\lambda)),
\ee
with
\be
  K(i/\lambda) = \min_{y \in [(i-1)/\lambda, i/\lambda[} K(y).
\ee
The continuous approximation of Eq.~(\ref{eq:1Dd})
\be
  \label{eq:1Dc}
  Q(h) \approx \exp\left(\lambda\int_0^h \log[F(K(y))]
  \mathrm{d}y\right),
\ee
holds for long distances as compared to the ``grain size''
$h\lambda\gg 1$ and a smooth stress intensity factor profile
$\lambda dK(y)/dy\ll K(y)$. This simple result was then extended
to the more complex case of subcritical crack growth~\citep{1322}.
However, this situation is not considered in the present study.

Let us note that a similar expression can be found by assuming
that the grain boundaries are defined by a Poisson
tessellation~\citep{1272}. Under this assumption, $1/\lambda$ is
the {\em average} grain size and the propagation probability with
a decaying stress intensity factor can be written
as~\citep{1272,1252}
\be
  \label{eq:1Ddj}
  Q(h) = \exp\left(-\lambda \int_0^h [1-F(K(y))]
  \mathrm{d}y\right).
\ee
%}
The key question that underlies this description is the
identification of the physical meaning of the ``grain size,'' and
the effective toughness.  In this context, the first sections that
considered the roughening of the crack front are helpful to
quantify more precisely these terms, and thus the domain of
validity of this description.  One of the major advantages of
resorting to this propagation model is that a well-defined
effective toughness is now available, naturally parameterized in
``time'' $t$, which is proportional to the mean crack advance
$t=\langle h\rangle_x L/(\xi_\parallel \xi_\perp)$ where
$\xi_\perp$ is the extension of an elementary crack advance during
one time step (\ie the rescaling of the length $L$ by
$\xi_\parallel$ is used to obtain the number of microstructural
units).

As noted above, the crack front develops a self-affine roughness,
thus the width of the front $w=(\langle h^2\rangle_x-\langle
h\rangle_x^2)^{1/2}$ is {\em dependent} on the crack length $L$,
obeying $w \propto \xi_\perp (L/\xi_\parallel)^\zeta$. Moreover,
complex correlations in the effective toughness are present.
However, those correlations only occur at time intervals smaller
than the time $T$ needed to depin the crack front over its entire
length. This defines the larger time scale for those correlations
scaling as $T \propto (L/\xi_\parallel)^z$. Taking into account
the fact that $L$ is present in the relationship between time and
mean crack advance, the correlation length $1/\lambda$ scales as
\be
  1/\lambda\propto \xi_\perp(L/\xi_\parallel)^{z-1}
  \propto \xi_\perp(L/\xi_\parallel)^{\zeta}.
\ee
Equivalently, the ``grain size'' $1/\lambda$ is physically given
by the width of the crack front. This signals a first non trivial
feature, unexpected from the one-dimensional picture. {\em The
``grain size,'' which has to be used, depends on the crack front
length}.

One arrives now at the most interesting output of the planar crack
modeling: namely, the distribution of effective toughness. The
power-law distribution of the effective toughness, which can be
seen in the vicinity of $\kappa^*$, has already been discussed in
Section~\ref{se:keff}. However, this power-law regime results from
the collective interaction of the crack front motion at different
scales.  Hence such a distribution holds in a domain where strong
correlations are present, and therefore it cannot be used in the
proposed approach given in Eqs.~(\ref{eq:1Dd}) and (\ref{eq:1Dc})
because of the necessary assumption of statistical independence.
However, the scale (\ie ``grain size'') at which the latter
condition holds is known. At such a scale, one has to consider the
maximum effective toughness over $1/\lambda$. Such a quantity has
already been used in the study of the effective toughness
conditioned by the distance between consecutive active sites, $d$.
It suffices to consider the largest distance $d=L/2$ admissible to
focus on the largest effective toughness. Moreover, the
distribution of effective toughness converges for large $d$ to a
universal scaling form given in Eq.~(\ref{eq:loaddis_d}). Thus the
shape of the effective toughness distribution $f$ is given by the
universal function $\psi$.  It simply has to be translated to
match $\kappa^*$ and scaled so that its standard deviation scales
as $(L/\xi_\parallel)^{-1/\nu}$. This is an important conclusion:
{\em When the crack length is large compared to the heterogeneity
of the microstructure, the effective toughness assumes a unique
universal form independent of the microstructure. However the
standard deviation of the distribution is scale-dependent.}

\begin{figure}
\centerline{\epsfxsize=.9\hsize \epsffile{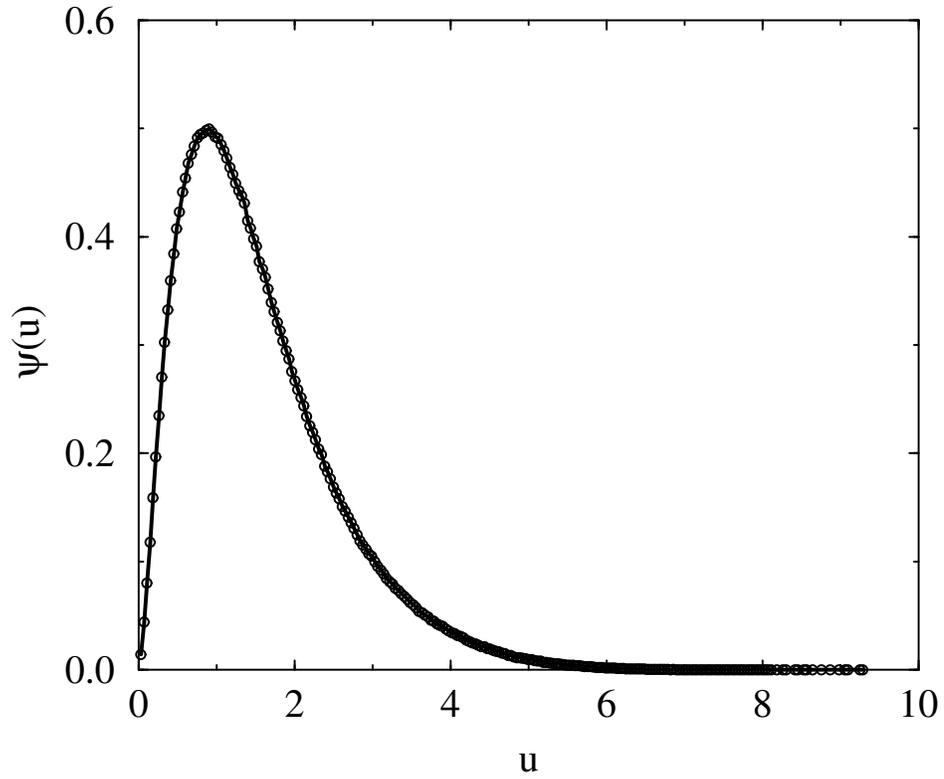}}
\caption{\label{fi:universal} Universal function $\psi(u)$ giving
the distribution of the effective toughness.  Note that the origin
$u=0$ corresponds to $\kappa=\kappa^*$, and increasing $u$
correspond to decreasing $\kappa$ values.}
\end{figure}

Figure~\ref{fi:universal} shows the shape of the universal
distribution $\psi$. This function has a power-law behavior close
to the origin, with the same exponent as the one of the global
toughness distribution, \ie $\psi(x) \propto x^\beta$. This
implies that the effective toughness does not exceed the threshold
$\kappa^*$.  For large arguments $x$, $\psi(x)$ has a rapid decay
to 0. This behavior concerns the small effective toughness and
turns out to be inessential as explicitly shown in the following
section.

\section{Application to indentation cracks}

\subsection{Propagation probability}

A natural application of the above discussion concerns indentation
cracks.  This situation is a typical case where the stress
intensity factor decreases strongly with the distance $c$ from the
indentation center, $K \propto M c^{-3/2}$, where $M$ is the
applied load~\citep{1241,1242}, so that cracks are generally
confined to the immediate vicinity of the indent. However,
radial/median cracks grow assuming approximately a circular or partly
circular shape~\citep{1431}.  Hence the effective crack front
length increases proportionally to $c$. This is a significant
difference with respect to the previously treated case where $L$
was assumed to remain constant.

The change of the crack front length with $c$ is responsible for a
continuous change of both the length parameter $1/\lambda$ and the
distribution of effective toughness. Coming back to a one
dimensional image, the problem consists now of a crack propagating
through a series of effective grains of increasing size, the
toughness of each grain being drawn from a distribution that is
itself grain size dependent. The generalization of
Eq.~(\ref{eq:1Dc}) leads to
\be
    \label{eq:2D}
    Q(c) \approx \exp\left[\int_0^c \log\left(F_r(K(r))\right)
    \lambda(r)\mathrm{d}r\right],
\ee
where the ``grain size'' $1/\lambda$ now depends on the crack
front length, $1/\lambda=B c$, with $B=\pi$ for a semi-circular
crack shape, $2\pi$ for a circular one.  The $B$ value is not
specified in the sequel to account for any crack shape provided it
remains self-similar at different stages of growth.  Moreover,
since the numerical prefactors in the scaling properties are not
given, their values can be disregarded at this stage
\be
    \frac{1}{\lambda(c)} \propto
    \xi_\perp\left(\frac{c}{\xi_\parallel}\right)^{\zeta},
\ee
% \SR{See above $\propto$ sign}
%
and the cumulative probability $F_r(K)$ of propagating through
that grain derives from a distribution whose shape is universal
(\ie independent on the details of the toughness disorder at the
microstructural scale, provided the latter is small compared with
the crack front length) but whose width directly depends on the
crack front length $L \propto c$
\be\ba{ll}
    F_c(K)&\displaystyle=1-\int_K^{K^*} f(k,c) \mathrm{d}k \\
    &\displaystyle=1-\int_0^\frac{K^*-K}{K^*}
     \left(\frac{c}{\xi_\parallel}\right)^\alpha
     \psi \left[q \left(\frac{c}{\xi_\parallel}\right)^\alpha\right]
    \mathrm{d}q.
\ea\ee
Note that for $K(c) > K^*$, the crack cannot be arrested so that
$F_c(K)=1$.  This leads to the introduction of a characteristic
crack size $c^*$ such that $K(c^*)=K^*$ which is a lower bound on
the crack size at arrest.  For a large crack size as compared to
the scale of heterogeneities, effective toughness values will
always remain in the immediate vicinity of $K^*$ (\ie $F$ close to
unity) so that the power-law behavior of $\psi$ at the origin can
be used to give a more precise description of $\log(F_c(K))$
\be
%    \log(F_c(K)) \propto -c^{\alpha(1+\beta)}(K^*-K)^{1+\beta}.
    \log(F_c(K)) \propto
-\frac{c}{\xi_\parallel}\left(\frac{K^*-K}{K^*}\right)^{\frac{1}{1-\zeta}}.
\ee
where the scaling relations summarized in Table~\ref{tab:expo}
have been used. Gathering these expressions, Eq.~(\ref{eq:2D}) can
be rewritten as
\be
 \label{eq:2Db}
 Q(c)\approx
 \exp\left\{ -A~(c^*)^{2-\zeta}
   \int_{c^*}^c
   \left(\frac{r}{c^*}\right)^{1-\zeta}
   \left(\frac{K^*-K(r)}{K^*}\right)^{\frac{1}{1-\zeta}}
 \frac{\mathrm{d}r}{c^*} \right\},
 \end{equation}
where $A$ is a  prefactor whose dependence with respect to the
material parameters $\xi$ is
  \be\label{eq:pref}
  A \propto \frac{\xi_\parallel^{\zeta-1}}{\xi_\perp}.
  \ee
A change of variable $x=c/c^*$ using the scaling of the stress
intensity factor $K(c)=K^* ({c^*}/{c})^{3/2}$ then leads to
%
%\begin{eqnarray}
\be
 \label{eq:2Dc}
 Q(xc^*)  \approx
 \exp\left\{ -A (c^*)^{2-\zeta}
 %\right. \nonumber \\ \times  \left.
    \int_1^{x}
   u^{1-\zeta} \left(1-u^{-3/2}\right)^{\frac{1}{1-\zeta}}
%   u^{\alpha(1+\beta)-\zeta} \left(1-u^{-3/2}\right)^{\beta+1}
   \mathrm{d}u \right\}.
\ee
%\end{eqnarray}
%
The integral in Eq.~(\ref{eq:2Dc}) can be recast in terms of an
incomplete beta function. It can be noted that the same
distribution was chosen {\em a priori} in \citep{1252}.
Equation~(\ref{eq:2Dc}) therefore constitutes an {\em a
posteriori} validation, even though the ``grain size'' is no
longer constant in the present analysis.

If the further hypothesis $(u-1) \ll 1$ can be made (\ie small
standard deviation of the crack arrest distance as compared to
$c^*$, in practice $x-1<5\%$), the following algebraic expression of
the decay of $Q(xc^*)$ with $x$ results
\be
% Q(xc^*) \approx \exp(-A' (c^*)^{2-\zeta} (x-1)^{(2-\zeta)/(1-\zeta)}),
 Q(xc^*) \approx \exp\left(-A' (c^*)^{2-\zeta} (x-1)^{\frac{2-\zeta}{1-\zeta}}\right),
\ee
where the numerical value of the above exponents is
$(2-\zeta)/(1-\zeta)\approx 2.6$ and $2-\zeta\approx 1.63$, and
the dependence of $A'$ on the characteristic parameters
$\xi_\perp$ and $\xi_\parallel$ is similar to that of $A$
[Eq.~(\ref{eq:pref})]. In particular, the width $\Delta$ of the
distribution of reduced crack size $c/c^*$ appears to scale as
$\Delta \propto (c^*)^{-(1-\zeta)} \approx (c^*)^{-0.63}$,
becoming more and more deterministic as the load is increased, \ie
$\Delta \propto M^{-2(1-\zeta)/3} \propto M^{-0.42}$, with a
power-law behavior reminiscent of a Weibull-type result, but with
the great difference that the exponent is {\em independent} of the
concerned material!

\subsection{Analysis of indentation experiments}

\begin{figure}
 \centerline{\epsfxsize=.9\hsize \epsffile{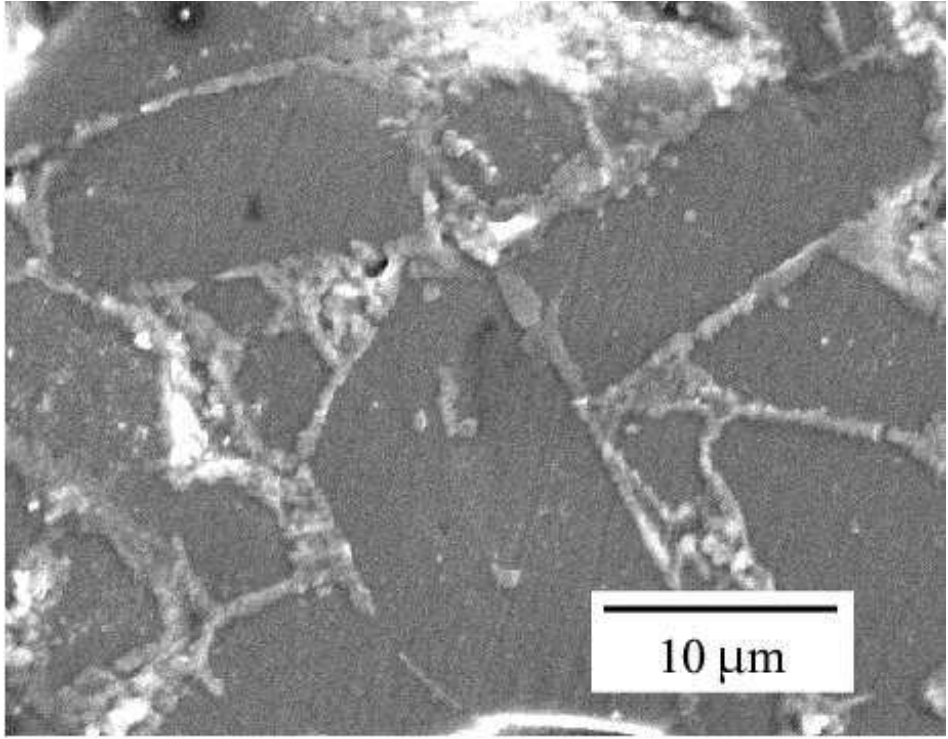}}
 \caption{\label{fi:Al2O3} Alumina used in the study with an
 average grain size of the order of 10~$\mu$m. One can note the
 glassy intergranular phase.}
\end{figure}

The previous results are now applied to experimental data. Several
indentation tests, with different applied masses $m$ (\ie $M
\propto m$), have been performed on an alumina (Al$_2$O$_3$)
ceramic. Figure~\ref{fi:Al2O3} shows an SEM picture of the
material. One can see a fine-grain (\ie 10$\mu$m) alumina
polycrystal with an intergranular glassy phase. A microanalysis
shows that the latter contains SiO$_2$, CaO and Al$_2$O$_3$
components. Vickers indentation with four different masses have
been performed on previously polished sample surfaces: namely,
$m=$ 0.2~kg, 0.3~kg, 0.5~kg and 1~kg. For each applied mass,
mean indentation-generated crack lengths are measured. For the
chosen loading range, it was checked that radial/median crack
system is predominant by SEM observations and that $c/2a >
1$, where $a$ is the size of the mean indent half-diagonal
length~\citep{1241,1242}. For each applied mass, measured crack
lengths $c_i$ are associated to an experimental propagation
probability $Q(c_i)$. The crack lengths are ranked in ascending
order (\ie $c_1 < c_2 < \ldots < c_N$) and the corresponding
experimental probability are evaluated as $Q(c_i) = 1 - i /
(N+1)$, where $N$ is the number of measured crack radii for the
considered applied mass.

\begin{figure}
\centerline{\epsfxsize=.9\hsize\epsffile{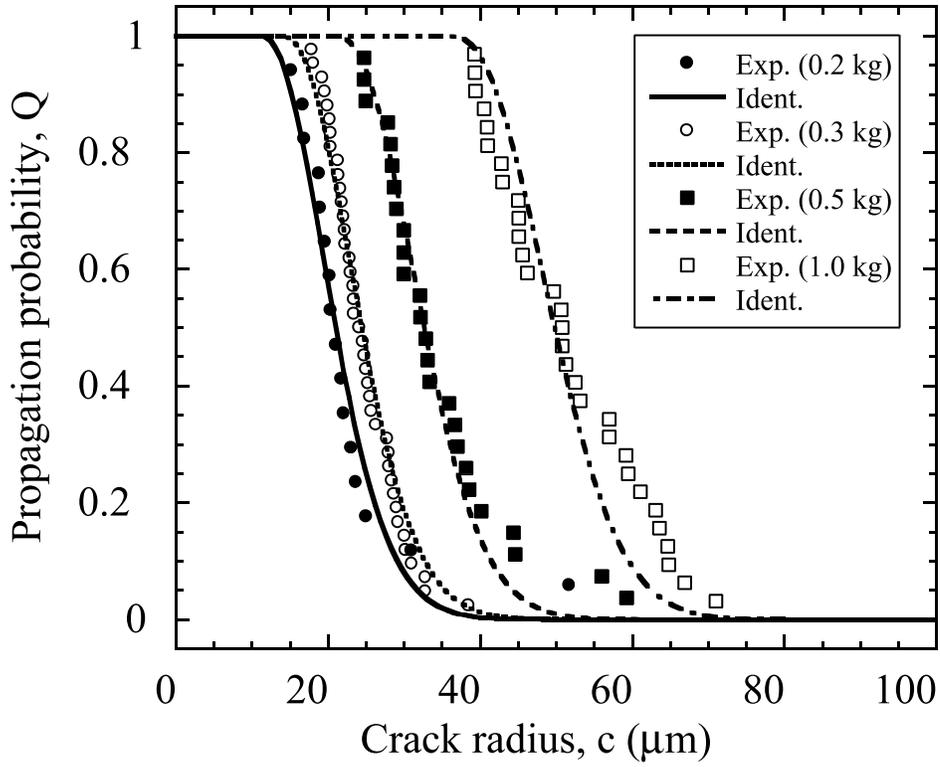}}
\caption{\label{fi:ident_bis_new}Propagation probability $Q$
versus crack radius $c$ for four different applied masses. The
symbols are experimental data and the lines are identifications
when the parameter $A$ is assumed to be load-independent.}
\end{figure}

Equation~(\ref{eq:2Dc}) is used in the following analysis. Only
two parameters have to be identified, namely the scale parameter
$A$ and the characteristic radius $c^*$.  Only one identification
procedure is presented here, although a number of alternative methods
can be used~\citep{1621}. Following Equation~(\ref{eq:2Dc}), for
each load series, the best $c^*$ parameter is sought by
prescribing that the constant $A$ be an intrinsic material
parameter (\ie unknown but identical for all series).
Figure~\ref{fi:ident_bis_new} shows the result of the
identification. A good agreement is obtained.
Table~\ref{tab:ident-ter} gives the value of the parameters $c^*$
for the different applied masses. The value of $A^{1/(2-\zeta)}$
is found to be equal to 0.15~$\mu$m$^{-1}$.

\begin{figure}
\centerline{\epsfxsize=.9\hsize\epsffile{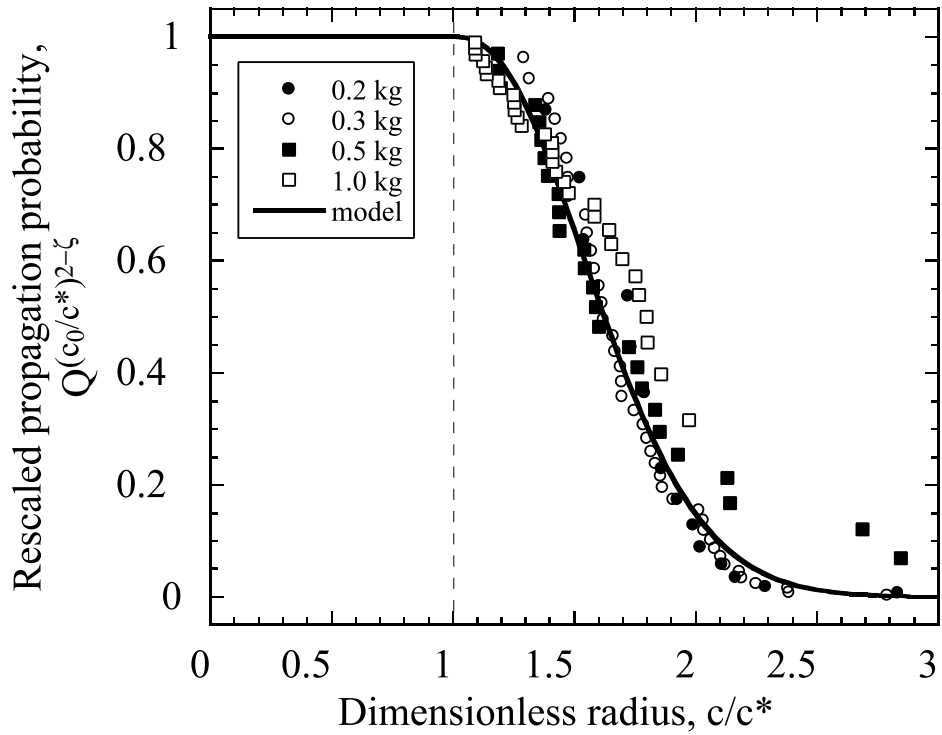}}
\caption{\label{fi:ident_bis_new_col}Rescaled propagation
probability $Q_{rescaled}$ versus dimensionless crack radius
$c/c^*$ for four different applied masses. The symbols are
experimental data and the curve is the result of the
identification with a single parameter $A$. From the present
analysis, it is expected that all experimental points should fall
on the same curve.}
\end{figure}

Furthermore, a rescaling procedure can be followed to compare all
the experimental data on a {\em single} curve.  Because of the
algebraic expression of $Q$, one has to refer to a common
conventional $c^*$, denoted as $c_0$, to be able to compare all
data on a unique scale.  Thus a {\em rescaled} propagation
probability, $Q_{rescaled}$, is introduced such that
 \be
 Q_{rescaled}=Q^q
 \ee
 where
\be
    q = \left(\frac{c_0}{c^*}\right)^{2-\zeta}.
\ee
A check of the proposed scaling result is that $Q_{rescaled}$
plotted as a function of $c/c^*$ should fall onto a unique master
curve.  This is expected to be true for any $c_0$ choice.  To
treat all experimental points on an equal footing, the
selection of the geometrical average of $c^*$ over all series was
made, $c_0=\exp(\langle \log(c^*_i)\rangle)$ where $\langle
...\rangle$ denotes the arithmetic average.
Figure~\ref{fi:ident_bis_new_col} shows the prediction for the
indentation experiments on alumina, and a good data collapse is
observed.

\begin{figure}
\centerline{\epsfxsize=.9\hsize \epsffile{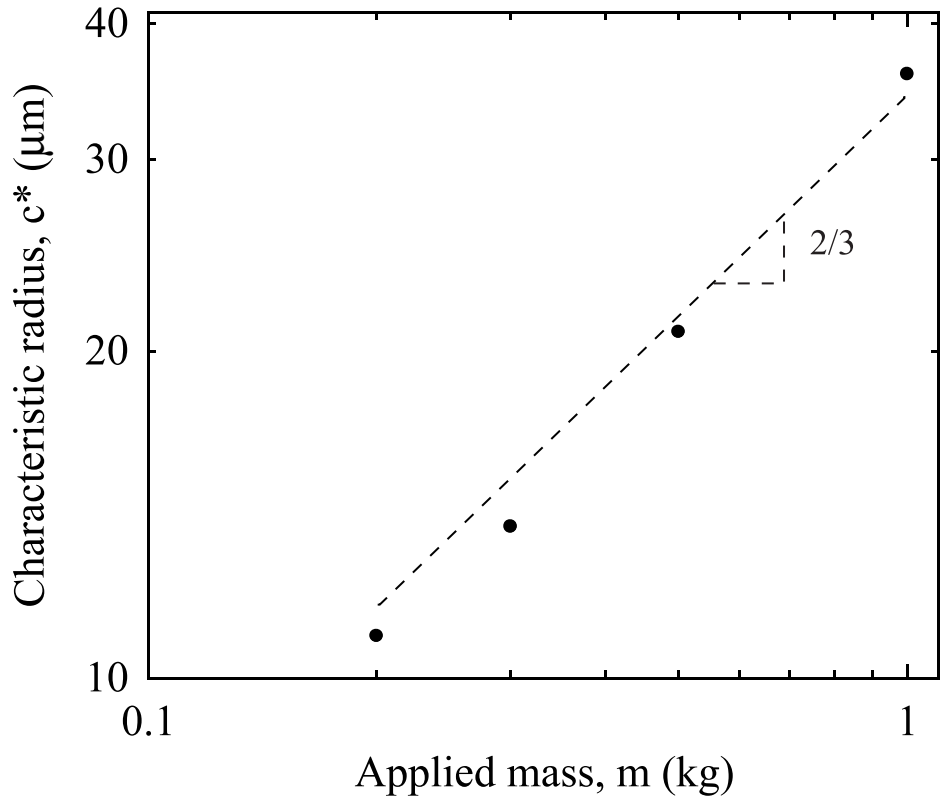}}
\caption{\label{fi:cstar-m} Parameter $c^*$ versus applied mass
$m$. A power law with an exponent of 2/3 fits reasonably well the
experiments (dashed line).}
\end{figure}

Finally, Equation~(\ref{eq:2Dc}) also predicts that $c^*$ follows
a power-law of the applied load $m$. Figure~\ref{fi:cstar-m} shows
this comparison in a log-log scale together with a line showing
the expected power-law dependence (\ie $c^* \propto m^{2/3}$
through the definition of a stress intensity factor $K^* \propto m
/ (r^*)^{3/2}$). Again a good agreement is obtained, although no
more fitting parameter was used in this last confrontation.

\section{Conclusions}

A model for the propagation of a planar crack in a two-dimensional
random landscape of toughness has been introduced. The complex
interplay between the disorder and the elastic crack does not
allow one to consider the successive failures of individual grains
as independent events. The onset of propagation can be described
as a genuine second order phase transition characterized by
universal critical exponents.

When the toughness fluctuations are large enough or when the
associated correlation length in the direction of propagation is
short enough, the advance of the front stops being steady.  The
motion becomes jerky, the front jumps from one stable position to
another one. In these strong pinning conditions the effective
toughness of the material cannot be estimated by a simple
arithmetic average of the toughness of the individual grains.
These successive bursts take place over a continuous range of
scales without any characteristic length besides the grain size
and the lateral extension $L$ of the front. Over these scales the
front develops universal features that do not depend on the
details of the toughness disorder. In particular, the front
roughness appears to be self-affine: its width scales as $w(L)
\propto L^\zeta$ with $\zeta\approx 0.37$. The motion of the front
is thus correlated up to the system size $L$ in the lateral
direction and $L^\zeta$ in the direction of propagation. When a
front of finite extent $L$ is considered, a coarse-grained
approach at the scale of effective grains of size $L\times
L^\zeta$ can be performed. A key result is that up to scaling
factors only dependent on the system size, the distribution of
effective toughness of these effective grains adopts a universal
form. The latter is completely independent of the details of the
microscopic toughness. If the lateral extent of the front remains
constant during the propagation, the latter can thus be treated as
the advance of a one-dimensional crack through a series of
effective grains characterized by the same well-defined effective
toughness p.d.f. This result is extended to the case of
self-similar cracks when the lateral front size is itself a
growing quantity in the course of propagation. The coarse-grained
approach remains valid but instead of advancing through effective
grains of constant size, the one-dimensional crack propagates
through effective grains of growing size together with a
non-steady toughness distribution.

As an application to explore the predictive power of this scaling
approach, the statistical distribution of crack arrest distances
for indentation has been examined. The proposed scaling approach
provides an analytical expression for the crack arrest length
distribution whose dependence with respect to the material
characteristics is limited to only two scalar parameters (a length
scale, and a reference toughness).  The rest of the analytical
expression is universal,  (independent of the considered
material). These results have been validated against experimental
data obtained from a polycrystalline sintered alumina.

\newpage

\listoftables

\newpage

\begin{table}
\caption{\label{tab:expo}Summary of the theoretical expressions
and numerical values of the critical exponents defined in the text
expressed as a function of $\zeta$.}
\vskip1cm

\begin{center}
\begin{tabular}{|c|c|c|}
\hline
Exponent &Expression &Value\\
\hline
$\zeta$  &                     &0.37\\
$\nu$    &$1/(1-\zeta)$        &1.59\\
$\alpha$ &$(1-\zeta)$          &0.63\\
$\beta$  &$\zeta/(1-\zeta)$    &0.59\\
$z$      &$(1+\zeta)$          &1.37\\
\hline
\end{tabular}
\end{center}
\end{table}

\null
\vskip1cm

\begin{center}
Charles et al.
\end{center}

\newpage

\begin{table}
\caption{\label{tab:ident-ter}Values of the parameters $c^*$ for
the four different applied masses $m$ when the parameter $A$ is
assumed to be load-independent.}

\vskip1cm

\begin{center}
\begin{tabular}{|c|c|}
\hline
Mass $m$ (kg) & $c^*$~($\mu$m) \\
\hline
 0.2 & 11 \\
 0.3 & 14 \\
 0.5 & 21 \\
 1.0 & 36 \\
\hline
\end{tabular}
\end{center}
\end{table}

\null
\vskip2cm

\begin{center}
Charles et al.
\end{center}

\newpage

\listoffigures

\newpage

\null
\vskip2cm

\begin{center}
Charles et al.
\end{center}

\newpage

\null
\vskip2cm

\begin{center}
Charles et al.
\end{center}

\newpage

\null
\vskip2cm

\begin{center}
Charles et al.
\end{center}

\newpage

\null
\vskip2cm

\begin{center}
Charles et al.
\end{center}

\newpage

\null
\vskip2cm

\begin{center}
Charles et al.
\end{center}

\newpage

\null \vskip2cm

\begin{center}
Charles et al.
\end{center}

\newpage

\null \vskip2cm

\begin{center}
Charles et al.
\end{center}

\newpage

\null
\vskip2cm

\begin{center}
Charles et al.
\end{center}

\end{document}